\pdfoutput=1
\documentclass[final,conference,twocolumn,a4paper]{IEEEtran}
\usepackage{booktabs}  
\usepackage{multirow} 

\usepackage{graphicx}
\usepackage[utf8]{inputenc}

\usepackage{enumitem}
\setlist[itemize]{leftmargin=*}

\usepackage{amssymb}
\graphicspath{{figs/}}

\usepackage{url} 
\usepackage{hyperref}  
\usepackage{wrapfig}
\usepackage{units}

\usepackage[printonlyused]{acronym}

\usepackage{calc}
\usepackage{amsmath}
\usepackage{amssymb}
\usepackage{amsfonts}
\usepackage{mathtools}

\usepackage{color, colortbl}
\usepackage{multirow} 
\usepackage[caption=false]{subfig}
\usepackage{stfloats}

\usepackage{cleveref}
\usepackage{listings}
\usepackage{tikz}
\usetikzlibrary{arrows,positioning,automata,shadows,fit,shapes}

\definecolor{mygreen}{RGB}{28,172,0} 
\definecolor{mylilas}{RGB}{170,55,241}

\definecolor{BgGray}{gray}{0.7}%
\definecolor{BgGray2}{gray}{0.96}%
\definecolor{RowColorOdd}{named}{BgGray2}%
\definecolor{RowColorEven}{named}{white}%
\definecolor{comments}{gray}{.5}
\definecolor{Gray}{gray}{0.85}

\usepackage{pifont}

\definecolor{amber}{rgb}{1.0, 0.75, 0.0}
	
\usepackage[normalem]{ulem} 

\newcommand{\methodname}{XZero}
\newcommand{\lte}{l}

\newcommand{\numLTEantenna}{K}

\newcommand{\delay}{\tau}
\newcommand{\delayBackhaul}{\delay_{b}}

\DeclareMathOperator*{\argmin}{arg\,min}

\newcommand{\leftq}{``}

\begin{document}
\bstctlcite{IEEEexample:BSTcontrol}

\title{\methodname: On Practical Cross-Technology Interference-Nulling for LTE-U/WiFi Coexistence} 
\author{
	\IEEEauthorblockN{Anatolij Zubow, Piotr Gawłowicz, and Suzan Bayhan}
	\IEEEauthorblockA{Technische Universit\"at Berlin, Germany, \{zubow, gawlowicz, bayhan\}@tkn.tu-berlin.de} 
}

\maketitle

\begin{abstract}

%
LTE-U/WiFi coexistence can be significantly improved by placing so-called coexistence gaps in space through cross-technology interference-nulling~(CTIN) from LTE-U BS towards WiFi nodes.
Such coordinated co-existence scheme requires, for the exchange of control messages, a cross-technology control channel~(CTC) between LTE-U and WiFi networks which was presented recently. 
However, it is unclear how a practical CTIN operates in the absence of channel state information which is needed for CTIN but cannot be obtained from the CTC.
%
%
%
We present \methodname, the first practical CTIN system that is able to quickly find the suitable precoding configuration used for interference nulling without having to search the whole space of angular orientations.
%
\methodname~performs a tree-based search to find the direction for the null beam(s) by exploiting the feedback received from the WiFi AP on the tested null directions.
We have implemented a prototype of \methodname~using SDR platform for LTE-U and commodity hardware for WiFi
and evaluated its performance in a large indoor testbed.
Evaluation results reveal on average a reduction by 15.7\,dB in interference-to-noise ratio at the nulled WiFi nodes when using a ULA with four antennas. Moreover, \methodname~has a sub-second reconfiguration delay which is up to $10\times$ smaller as compared to naive exhaustive linear search. 

\end{abstract}

 
\section{Introduction}\label{sec:introduction}

As LTE operators have been expanding their operation to unlicensed spectrum via carrier aggregation, ensuring coexistence in the unlicensed bands has become a big challenge, particularly with WiFi networks which carry a significant fraction of current mobile data traffic and thereby are vital to communications~\cite{cisco0217}.
A significant body of research, e.g., \cite{olbrichwiplus,sagari2015coordinated}, develops solutions for this key issue and the common approach is to employ so-called \textit{coexistence gaps} in either time, frequency, space, or code domains. 
Coexistence gap can be defined as a resource block~(in one of the above-listed domains) which one network leaves to the other's use.   
LTE-U applies duty-cycling to fairly share the medium with the colocated WiFi networks~\cite{zhang2017ltesurvey}, i.e., uses time domain coexistence gaps.

In our recent work~\cite{bayhan2017coexistence}, we showed that LTE-U/WiFi coexistence can be significantly improved by placing coexistence gaps in space through cross-technology interference-nulling~(CTIN) from LTE-U BS towards WiFi nodes. 
Nulling removes the co-channel interference from LTE-U towards the nulled WiFi nodes; hence co-existence can be significantly improved. 
Consequently, the LTE-U network is able to use a larger duty-cycle in comparison to the usual one computed according to LTE-U's CSAT algorithm as the nulled WiFi node is effectively being removed from the system. 
But, using some of its degrees of freedom for nulling, the SNR towards its own UEs is slightly decreased. 
However, as long as the gain from increased CSAT cycle outweighs the SNR loss, nulling improves the capacity of LTE cell without deteriorating that of the WiFi cell. 
Moreover, under certain circumstances the sum/total capacity of the two networks can be increased~\cite{bayhan2017coexistence}.


\begin{figure}[!t]
\centering
   \includegraphics[width=0.85\linewidth]{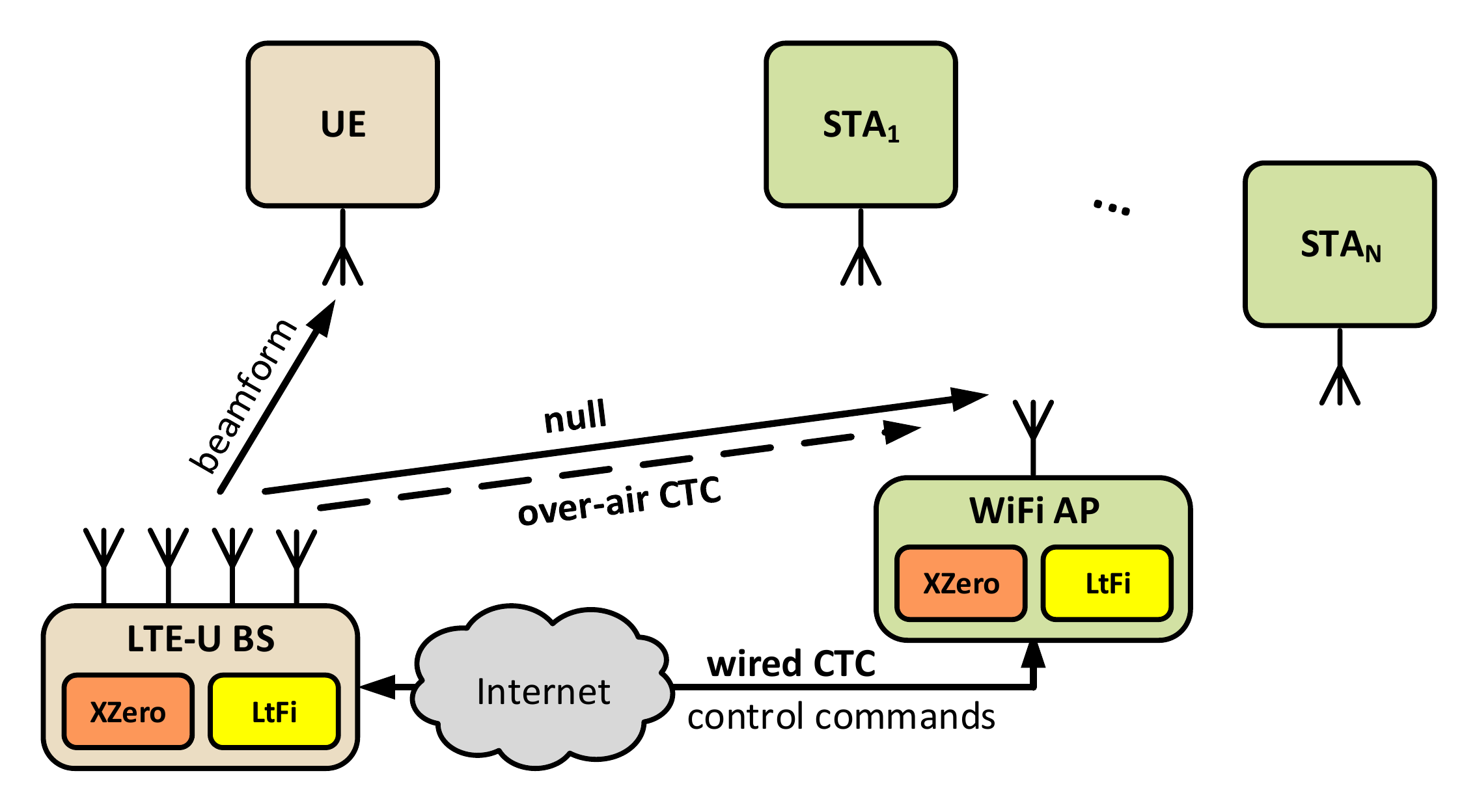}
 \vspace{-10pt}
\caption{Architecture of \methodname: LTE-U BS performs cross-technology interference-nulling towards co-located WiFi nodes. Signaling information is exchanged over cross-technology control channel (CTC).}
\label{fig:system_model}
\vspace{-15pt}
\end{figure}

Despite its promises, CTIN has its own challenges to be realized in a practical system. First, such coordinated co-existence scheme requires a cross-technology control channel~(CTC) between LTE-U and WiFi which is needed to exchange CTIN control information between both networks. Unfortunately such a CTC is missing in the standard WiFi/LTE implementations. Luckily, a recent work~\cite{ltfi_infocom} designs a CTC, named LtFi, which facilitates communication between LTE and WiFi networks. 
Unfortunately, the channel state information~(CSI) at the LTE-U BS towards the WiFi node to be nulled is needed for CTIN but it cannot be obtained from the CTC as only the amplitude of the signal is available.
%

Previous works overcame this challenge~(i.e., lack of CSI for nulling) by exploiting channel reciprocity for estimating the interferer’s channel ratio~\cite{gollakota2011clearing} which is sufficient to compute nulling precoding vector.
%
%
However, nulling in TIMO~\cite{gollakota2011clearing} is possible only if the colocated technology has a  bidirectional communication as channel reciprocity is exploited to estimate the channel ratios and it needs time for channel estimation in the order of a few seconds.
It remains unclear how TIMO's approach of utilizing channel ratios can be generalized to more than two antennas and multiple user nulling without requiring a high overhead~\cite{lakshmanan2009practical}.
In contrast, we design in the present paper \methodname~which does not exhibit these shortcomings as LTE-U BS does not estimate channel ratios but rather relies on a null search using the feedback from the WiFi nodes to be nulled.

%

%


\medskip

\noindent \textbf{Contributions:} We present \methodname, the first low-complexity solution enabling cross-technology interference nulling from LTE-U towards WiFi nodes. 
\methodname~is able to quickly, e.g., in sub-seconds, find a proper precoding configuration used for interference nulling without having to  search  the  whole  angular  space.  Instead  of  an  exhaustive linear null search, \methodname~performs a tree-based search to find proper  beamforming  configuration  for  nulling  the  preferred WiFi nodes.
We implemented a prototype of \methodname~and evaluated its performance in a large indoor testbed.
Our experiments show that \methodname~is able to decrease the interference-to-the-noise~ratio~(INR) at the WiFi nodes drastically, e.g., on the average 15.7\,dB. 
Moreover, \methodname~is very fast compared to a naive linear search which tests sequentially each candidate nulling configuration.
Another merit of \methodname~is its UE-transparent operation: LTE BS can simultaneously continue its downlink transmission while searching for the nulling angle.


%
%
%

%
%
\section{Background on LTE-U and LtFi}
\subsection{LTE-U/WiFi Coexistence Primer}\label{lteu_primer}


LTE-U is being defined by the LTE-U forum~\cite{lteu_forum} as the first cellular solution operating in the unlicensed bands for the downlink~(DL) traffic. The LTE carrier aggregation framework supports utilization of the unlicensed band as a secondary cell in addition to the licensed anchor serving as the primary cell. The LTE-U channel bandwidth is set to 20\,MHz which corresponds to the smallest channel width in WiFi.
LTE-U enables coexistence with WiFi by means of dynamic channel selection and adaptive duty cycling. For the latter, a mechanism called \textit{carrier sense adaptive transmission} (CSAT) is used to adapt the duty cycle, by modifying the $T_{\mathit{on}}$, and $T_{\mathit{off}}$ values where $T_{\mathit{csat}} = {T_{\mathit{on}}+T_{\mathit{off}}}$, to achieve fair sharing with WiFi and other LTE-U networks. 
More specifically, LTE-U adapts its $T_{\mathit{on}}$ value according to the observed WiFi medium utilization
and number of WiFi nodes~\cite{lteu_qualcom2016}.
Note that the LTE-U transmissions take place only during the $T_{on}$ phase. This is different to WiFi which applies a listen-before-talk (LBT) scheme.
According to \cite{lteu_qualcom2016}, a CSAT period $T_{csat}$ of 40, 80, or 160\,ms is recommended. 
Finally, the LTE-U transmissions contain frequent gaps during the $T_{on}$ phase (so called subframe puncturing), which allow WiFi to transmit delay-sensitive data. At least 2\,ms puncturing has to be applied every 20\,ms according to Qualcomm's proposal~\cite{lteu_qualcom2016}. 
Please refer to \cite{zhang2017ltesurvey} for more details on LTE in the unlicensed spectrum.


\subsection{LtFi Primer}\label{ltfi_primer}


LtFi~\cite{gawlowicz2017ltfi} is a system which enables to set-up a cross-technology control channel~(CTC) between co-located LTE-U and WiFi networks for the purpose of cross-technology collaboration, e.g., radio resource and interference management. 
It is fully compliant with LTE-U technology, and works with WiFi commodity hardware by utilizing the spectrum scanning capability of modern WiFi NICs~(e.g. Atheros 802.11n/ac). 
The LtFi architecture consists of two parts, namely an air and a wired interface. 
The first is used for over-the-air broadcast transmission of configuration parameters (i.e. IP address) from LTE-U BSs to co-located WiFi APs which decode this information by utilizing their spectrum scanning capabilities. 
This configuration data is needed for the subsequent step to set-up a bi-directional control channel between the WiFi nodes and the corresponding LTE-U BSs over the wired backhaul, e.g. Internet.
Note that a WiFi node, i.e., LtFi receiver, is able to measure on its air-interface the signal power of the LTE-U signal for each WiFi OFDM subcarrier $|h_i|^2$ during both LTE-U's $T_{on}$ and $T_{off}$ phase. 
Please refer to \cite{gawlowicz2017ltfi} for more details on LtFi.


\section{Illustrative Example}\label{sec:example}

We consider a coexistence setting as in Fig.~\ref{fig:system_model} which consists of an LTE-U network and a WiFi network both operating in the same unlicensed spectrum.
In this setting, our goal is to enable LTE-U BS to apply beamforming towards its users while nulling its interference on selected WiFi nodes~\footnote{Note, we assume that the LTE-U BS has already decided on which nodes to null and we concentrate on the actual step of nulling. Please refer to \cite{bayhan2017coexistence} for more on how the nodes can be selected for nulling.}.
Before going into details, let us explain \methodname~using an example.

\begin{figure*}[htb]
\centering
  \begin{tabular}{@{}c@{}}
    \includegraphics[width=0.9\textwidth]{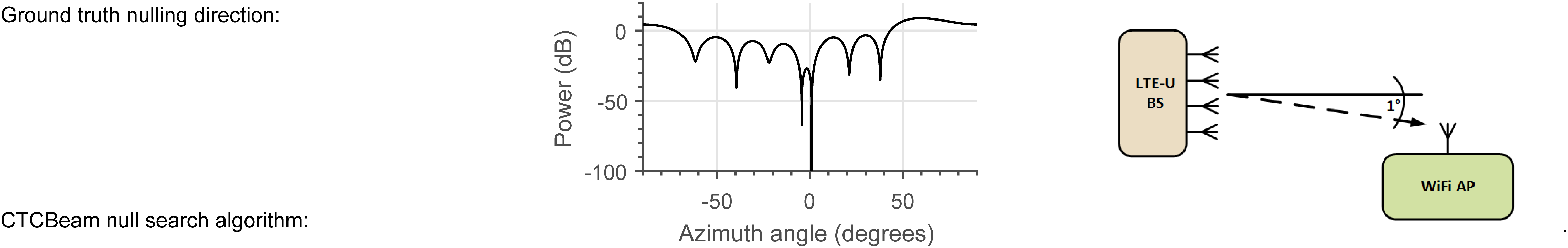}   \\
    \includegraphics[width=0.9\textwidth]{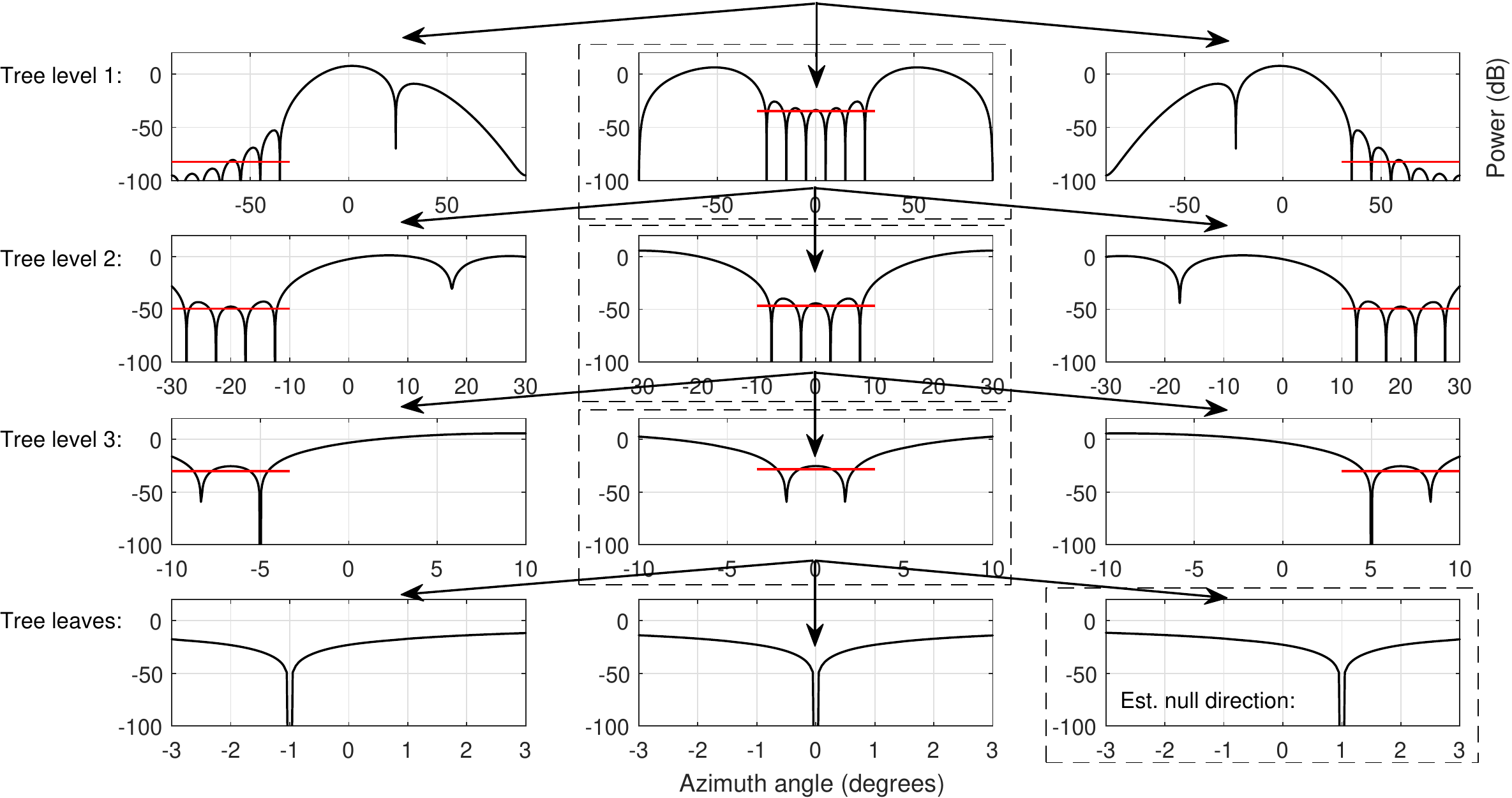}
  \end{tabular}
  \caption{Illustrative example of \methodname~null search algorithm (K=8 antennas). The WiFi AP is to be nulled and it is located at 1$^{\circ}$ from the LTE-U BS.
  }
  \vspace{-10pt}
  \label{fig:example} 
\end{figure*}

The major challenge of applying cross-technology interference nulling from LTE-U towards WiFi is the missing complex channel state information (CSI) which cannot be obtained from the LtFi CTC air-interface. 
%
%
Hence, \methodname~employs an indirect approach using a random nulling scheme together with a tree search. 
Fig.~\ref{fig:example} shows how \methodname~estimates the nulling direction (1$^{\circ}$ in this example) by performing a tree-based search where the LTE-U BS is equipped with Uniform Linear Array~(ULA) with $K=8$ antennas.
It executes null sweeps in multiple steps. Specifically, it starts with a \leftq wide" null (beam) pattern to do a tree search and narrow the beam in subsequent steps.
In case of symmetric ULA, we break at level 1 the range from -90$^{\circ}$  to 90$^{\circ}$  into three sectors.
Each sector is again split up in smaller sectors at the next level, e.g. at level 2 the range from -30$^{\circ}$  to 30$^{\circ}$  is split up into three sectors.
Note that the number of nulls at each level are different, i.e. 6, 4, 2 and 1. This is needed to guide the null-beam search in the right subtree, i.e., there is exactly one sector giving the lowest interference-to-noise ratio~(INR) at each tree level. 
A null beam configuration is kept for the duration of a single LTE-U cycle period, $T_{\mathit{csat}}$.
After testing all nulling configurations at a particular level, the WiFi node reports the sector having the lowest INR to the LTE-U BS through LtFi's wired interface.
Next, BS continues its null search in the subtree of the reported nulling configuration.
%
The algorithm stops after testing all leaf nodes. Note that leaf nodes have just a single null.
Our tree-based search is similar to~\cite{hunzinger2006intelligent} except that we search for null beams.

\section{System Model}

We consider a system consisting of a single LTE-U BS and a colocated WiFi BSS~(AP with multiple STAs) as depicted in Fig.~\ref{fig:system_model}.
BS and AP are connected to a wired backhaul and are able to exchange messages over the backhaul with a delay of $\delayBackhaul$\,ms.
Both the LTE-U and the WiFi cell operate on the same unlicensed channel with bandwidth of 20\,MHz. 
The BS is equipped with a Uniform Linear Array (ULA) with $\numLTEantenna$ antenna elements. The AP is equipped with a single antenna.
The transmit power budget for the BS is $P_{\lte}$, i.e., transmit power after precoding sums up to $P_{\lte}$.
For the BS, we consider the DL case where it adjusts the transmitted signal by means of precoding in order to achieve interference nulling. The precoding weights can be assigned on a per Radio Resource Block (RRB) level. The total number of RRBs is $\mathrm{NRRB}$.
%
LTE-U applies adaptive duty cycling as described in Section~\ref{lteu_primer}. 
%
%
%
%
As mentioned in Section~\ref{ltfi_primer}, the AP is able to measure the received signal power of frames sent over the CTC air-interface on a per WiFi OFDM subcarrier granularity, i.e., $|h_s|^2, s=1 \ldots \mathrm{NSC}$, where $\mathrm{NSC}$ is the number of subcarriers.

\section{\methodname's Architecture}\label{sec:approach}

\subsection{Overview of \methodname}

The main objective of \methodname~is to improve the co-existence between co-located LTE-U and WiFi networks. 
To this end, it performs at the LTE-U BS interference nulling towards co-located WiFi nodes (Fig.~\ref{fig:system_model}). As a consequence, the BS is able to increase its CSAT duty cycle~\cite{bayhan2017coexistence}.
Another goal is to make sure that the performance of both networks is not significantly deteriorated during the null search. 
Hence, instead of searching with a beam~(e.g., in contrast to ~\cite{hunzinger2006intelligent}) we are searching with nulls to ensure that WiFi is not affected. 
To avoid degradation in LTE network throughput during the null search, the signal towards the LTE UE is always beamformed~(cf.~\ref{fig:system_model}).
Hence, an LTE-UE can continue to receive its DL traffic while BS is null searching.
\methodname~requires no modifications to the WiFi besides the support of LtFi CTC. 
On the BS side, we assume the existence of an SDR platform which gives the possibility to perform antenna precoding.
All required signaling information is exchanged using the LtFi system.



%
%
%
\subsection{Precoding Vectors}\label{sec:precoding_vec}

As shown in the example from Section~\ref{sec:example}, the BS performs a tree-based null search to find the best nulling configuration. In \methodname, the precoding vector is computed using LCMV beamformer~\cite{van2002optimum} as it allows us to put the signal in the desired direction~(i.e., UE) and to place nulls into other directions, i.e., WiFi nodes. The inputs to the LCMV are the direction of arrival angles.
In \methodname, the precoding weight vectors $w \in \mathbb{C}^{1 \times K}$ are precomputed and stored in a tree data structure. During the null-search the tree is traversed.

\subsection{Power Correction}\label{sec:power_corr}

In free space environment without multipath reflections, the so far described approach is able to find the correct nulling angle~(for each LTE-U RB/SC), i.e., good INR values after nulling. However, this is not the case in a real  environment with significant multipath resulting in frequency-selective fading. This is because so far we do not take the geometry of the environment into account. 

Hence, before performing the actual null-search, we measure the power on each antenna path independently. Therefore, the BS is transmitting its signal on each transmitter antenna alternately.
The WiFi node to be nulled estimates the receive power $|h^k_s|^2$ of each antenna path $k$ on each WiFi OFDM subcarrier $s$. This information is feedbacked to the LTE-U BS which is using it to correct the precoding values so that the power in each antenna path stays the same.
However, the difference between the WiFi and LTE PHY layer, i.e., subcarrier and RB orientation, poses a challenge for \methodname~in this step.
In WiFi, each 20 MHz channel accommodates 64 subcarriers each with 312.5 kHz bandwidth whereas an LTE channel with 20 MHz bandwidth consists of RBs with 180 kHz bandwidth and  15 kHz subcarriers. 
To have a mapping between the measured signal at the WiFi receiver and LTE transmitter RB, we find the subcarrier $\hat{s}$ 
that has the closest central frequency to that of the LTE RB $r$, i.e., $\hat{s} = \argmin_{s \in \mathrm{NSC}}{|f_c(r) - f_c(s)|}$ where  $f_c(\cdot)$ gives the center frequency of a WiFi subcarrier or RRB. 
Note that one could apply other ways to achieve a more accurate estimation, e.g., extrapolation from 312.5 kHz to 180 kHz values. However, this aspect is out of scope for the present paper.

Let $W$ denote the actual BS's precoding weight matrix: $W \in \mathbb{C}^{K \times \mathrm{NRRB}}$, where $\mathrm{NRRB}$ is the total number of LTE RRBs. 
Then, we calculate the column $r$ corresponding to RRB $r$ of the corrected weight matrix as follows:
The $r^{\textrm{th}}$ column representing the precoding for RRB $r$ is computed as: 
\begin{align}
	W_r {=} w \odot \left(\frac{|h^0_{s}|^2}{|h_{s}|^2}\right)^\frac{1}{2} \text{, where } s{=}\argmin_{s \in \mathrm{NSC}}{|f_c(r) {-} f_c(s)|} \label{eq:W}
\end{align}
\noindent where $w$ is the precomputed precoding vector (Sec.~\ref{sec:precoding_vec}) and $\odot$ being the element-wise multiplication. 

In a final step, we normalize to ensure power after precoding sums up to the transmit power budget: \begin{align}
	W_r^* = \frac{W_r}{\| W_r \|_F} \label{eq:W_norm}
\end{align}
\noindent where $\| \cdot \|_F$ denotes the Frobenius norm.

\subsection{Standard Mode of Operation}\label{sec:ctcbeam_smo}

\begin{figure}[!t]
\centering
\begin{minipage}[b]{1\linewidth}
\begin{center}
   \includegraphics[width=1\linewidth]{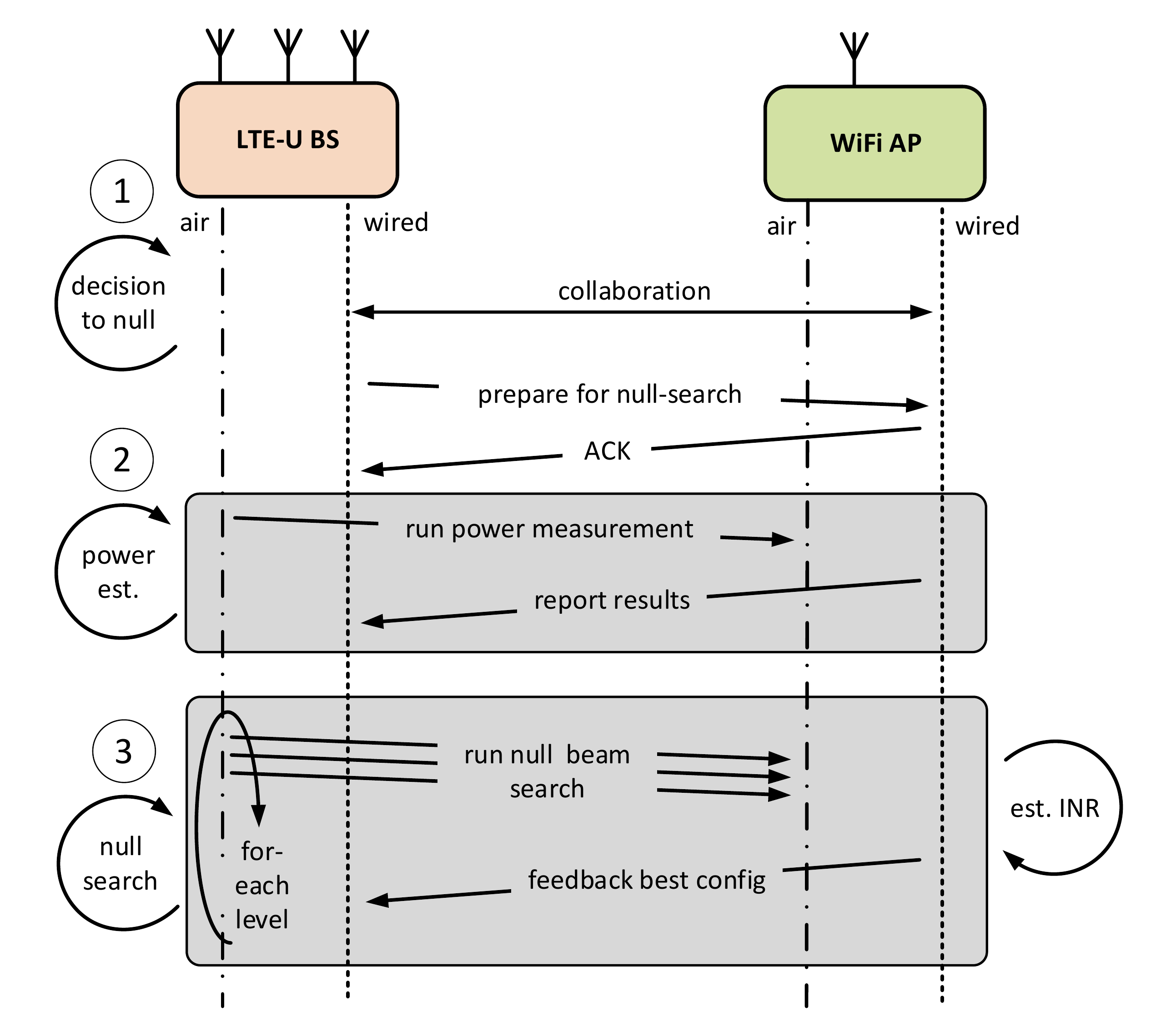}
\end{center}
 \vspace{-10pt}
\caption{The LTE-U and WiFi networks are collaborating with each other over the LtFi wired control channel. The process starts when the decision to null a particular WiFi node is made. Before the actual null-search, there is a phase where the receive power on each antenna path is measured.}
\label{fig:signalling}
\end{minipage}
\vspace{-5pt}
\end{figure}

Fig.~\ref{fig:signalling} shows the standard operation mode in \methodname. The LTE-U and WiFi networks collaborate with each other over the LtFi wired control channel. In case the decision was made to null the WiFi node, the power measurement phase starts at the end of which the BS knows the power on each antenna path and hence is able to correct the precomputed precoding values as described in Sec.~\ref{sec:power_corr}. 
Note that during that phase no precoding is applied.
The subsequent step is the actual tree-based null beam search during which the BS performs different nulling configurations for which the WiFi AP feedbacks the ID of the one having the lowest INR value. The search stops after testing the single null configurations, i.e., leaves. Finally, from all tested nulling configurations the one achieving the lowest INR value is used.

\subsection{Null Beam Search for Multiple Stations}\label{sub:multiple_users}

\begin{figure}[!t]
\centering
\begin{minipage}[b]{0.95\linewidth}
\begin{center}
    \includegraphics[width=1\linewidth]{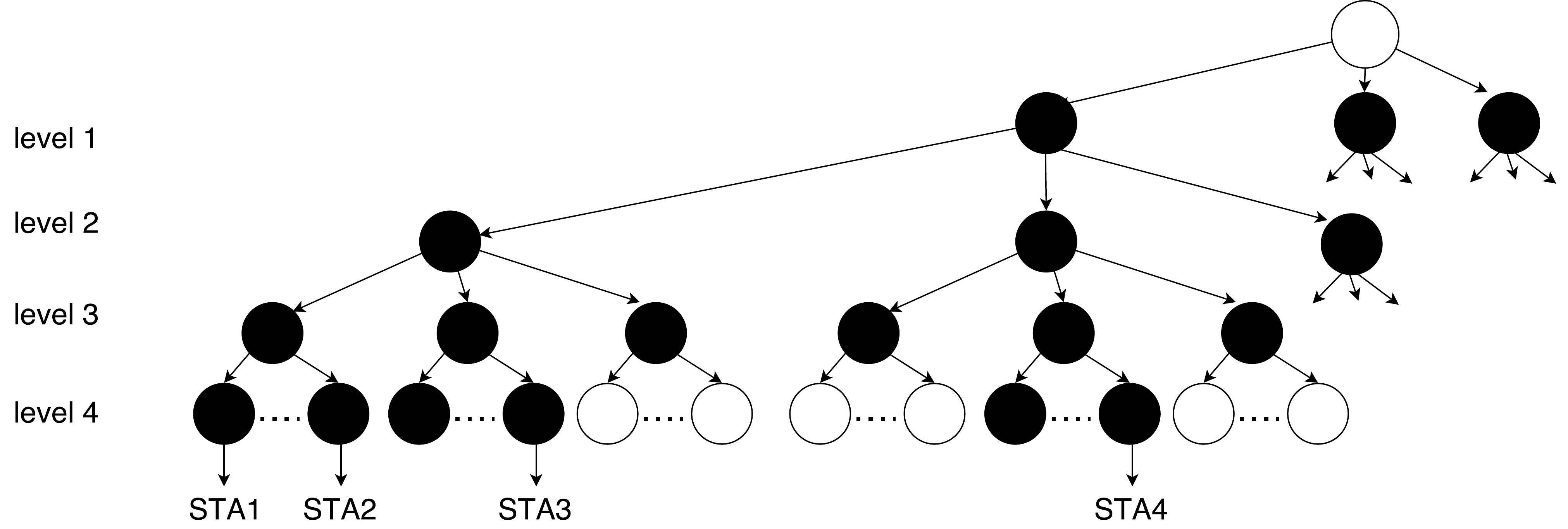}
\end{center}
 \vspace{-10pt}
\caption{Searching multiple users. Only the black nodes are visited.}
\label{fig:search_tree_Ntx8_parallel}
\end{minipage}
\vspace{-15pt}
\end{figure}

In case there are multiple users to be nulled at the WiFi cell, \methodname~can search in parallel to incur a lower reconfiguration delay compared to sequential search for multiple users. Fig.~\ref{fig:search_tree_Ntx8_parallel} shows how the tree search is executed for nulling four stations--- STA1 to STA4.
In this example, STA1 and STA2 are in a similar angular location relative to the LTE-BS and STA-4 is at a separate angular orientation compared to the remaining three stations.
Here, we can see that multiple subtrees are expanded in parallel. 
This is especially beneficial in case multiple WiFi nodes are co-located, i.e., at the same angular direction, where \methodname~achieves faster configuration than sequential execution of tree-search for each user.
%
Unfortunately, in the multi-user searching case the proposed power correction (Sec.~\ref{sec:power_corr}) cannot be applied resulting in slight degradation in INR per user (Sec.~\ref{sec:eval}).

\subsection{Extensions to LtFi}

For the purpose of \methodname~additional frame types were introduced in LtFi: i) power measurement and ii) null-beam search. The former is sent by the BS in preparation of the actual null-beam search to measure the power on each antenna path so that the precomputed precoding vectors can be corrected as suggested in Sec.~\ref{sec:power_corr}. The null-beam search frame marks the start of the tree-based during which different null-beam configurations are tested. 

\section{Implementation Details}\label{sec:proto}

We present details of our \methodname~prototype implementation.

\medskip

\noindent \textbf{LTE-U BS:}
The LTE-U BS was implemented on Ubuntu 16.04 LTS using srsLTE~\cite{srsLte}, the open-source software-based LTE stack implementation, running on top of USRP software-defined radio platform, namely X310.\footnote{\url{https://kb.ettus.com/X300/X310}} 
In particular, we modified srsLTE to implement LTE-U's duty-cycled channel access scheme, where we provide an API to program the duration of $T_{\mathit{on}}$ and $T_{\mathit{off}}$ of single LTE-U period as well as the relative position of the puncturing during $T_{\mathit{on}}$ phase. 
Moreover, the API allows setting the antenna precoding per RRB to be used during the LTE-U's $T_{\mathit{on}}$ phase in real-time using UniFlex~\cite{gawlowicz2017uniflex} control framework.
The actual functionality of \methodname~, i.e., the tree-based null search, and LtFi were implemented as Python-based applications.

\medskip

\noindent \textbf{WiFi AP:}
At the WiFi side, we use Ubuntu 16.04 LTS and commodity hardware, namely Atheros AR928X wireless NIC, that allows spectrum scanning at a very fine granularity.\footnote{\url{https://wireless.wiki.kernel.org/en/users/drivers/ath9k/spectral_scan}}
We sample with frequency of 5-50\,kHz~\footnote{Note that we used the maximum possible sample rate which is a chipset-specific value.} and pass this data to LtFi receiver component which is implemented entirely in Python.
Note that we disabled Atheros Adaptive Noise Immunity~(ANI).
%
Moreover, the LtFi receiver component reports the INR values measured during the LTE-U's $T_{\mathit{on}}$ and $T_{\mathit{off}}$ phases to the \methodname~component. From the set of measured INR values the \methodname~component estimates the nulling configuration with minimum INR which is feedbacked to the \methodname~component at the LTE-U BS side through the wired LtFi interface.
Finally, regarding the beamforming/nulling, no changes were needed on the WiFi side as the interference nulling is fully transparent to the receiver.

\section{Performance Evaluation}\label{sec:eval}

We evaluate the performance of \methodname~by means of experiments.
As performance metric, we report INR at the WiFi node, with and without nulling.
We calculate the INR as the ratio between the received power during the LTE-U on ~($P_{T_{\mathit{on}}}$) and off ~($P_{T_{\mathit{off}}}$) phases:
\begin{align}\label{eq:SINR}
\mathrm{INR} = \frac{P_{T_{\mathit{on}}}}{P_{T_{\mathit{off}}}},
\end{align}
where  $P_{T_{\mathit{on}}}$ is the interference from the DL LTE-U signal and $P_{T_{\mathit{off}}}$ corresponds to the noise in the environment as no LTE signal is transmitted during the off-period.
Subsequently, we calculate the interference reduction due to nulling as $\Delta$INR.

In the following, we have two sets of experiments: (i) over-the-cable experiments in our lab to mimic a controlled wireless channel where there are no multipath effects and (ii) over-the-air experiments in the ORBIT grid testbed representing a realistic wireless scenario.
Our aim is to understand the performance difference between \methodname's tree and linear search in terms of reconfiguration delay, i.e., due to mobility or joining/leaving of WiFi nodes, and the achieved $\Delta$INR at the nulled WiFi nodes. Moreover, we compare \methodname~with a baseline without interference nulling.

\subsection{Over cable experiments}

\begin{figure}[!t]
\centering
  \begin{tabular}{@{}c@{}}
    \includegraphics[width=0.85\linewidth]{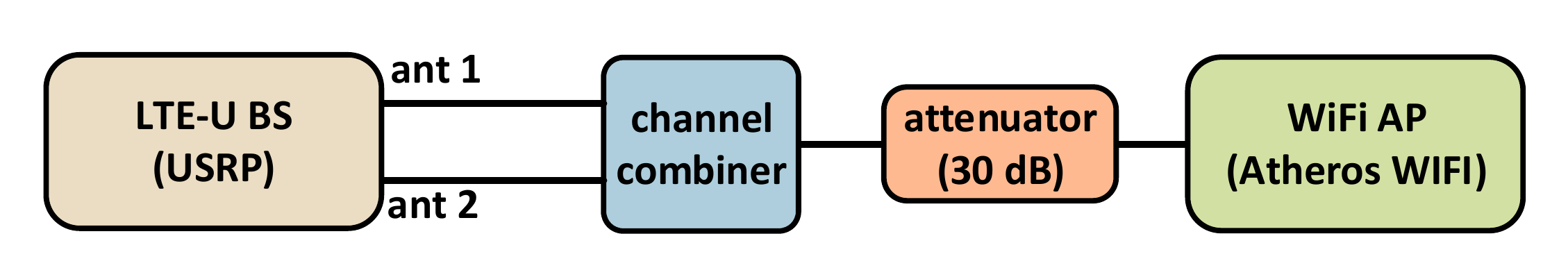}   \\
    \includegraphics[width=0.9\linewidth]{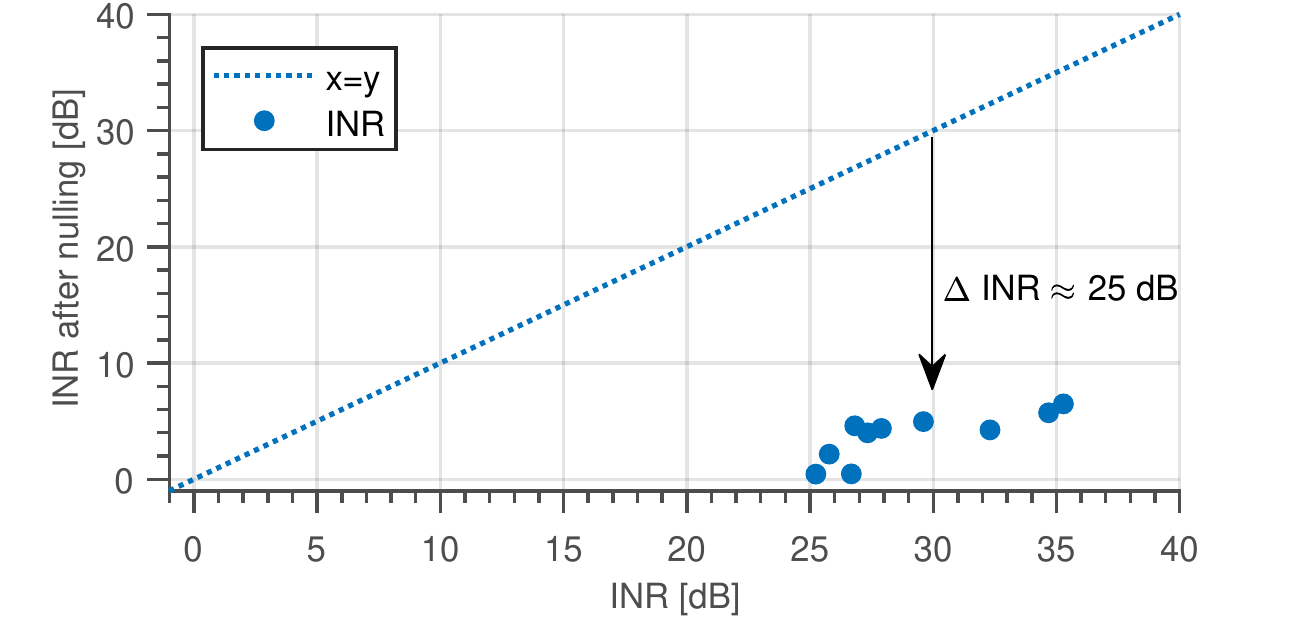}
  \end{tabular}
  \caption{Over cable experiment --- experiment setup (upper) and INR before vs. after nulling for different USRP transmit power level, i.e. 0 to 27\,dB, (lower).}
  \label{fig:cable_exp_inr} 
  \vspace{-10pt}
\end{figure}

In this scenario, we aim to understand the upper bound in $\Delta$INR \methodname~can provide under optimal conditions.
Hence, we use a wired (i.e. frequency flat) channel. We mix the output of BS's $K{=}2$ antenna ports using a channel combiner (Fig.~\ref{fig:cable_exp_inr}, upper).
Next, the combined signal was transmitted over coaxial cable to single WiFi node and 
the INR was measured.
Since the noise floor depends on the BS's (USRP) transmit gain, it was measured each time during the LTE-U off-phase.

Fig.~\ref{fig:cable_exp_inr} plots the measured INR values before and after nulling for different transmission power levels.
As we see from the figure, the measurement points are clustered below the $x=y$ line and close to x-axis, meaning that the INR before nulling is significantly higher than the INR after \methodname's nulling operation. On average, the interference is reduced by $\Delta$INR=25\,dB.

\subsection{Over-the-air experiment in ORBIT testbed}

\begin{figure}[!t]
\centering
\begin{minipage}[b]{1\linewidth}
\begin{center}
    \includegraphics[width=1\linewidth]{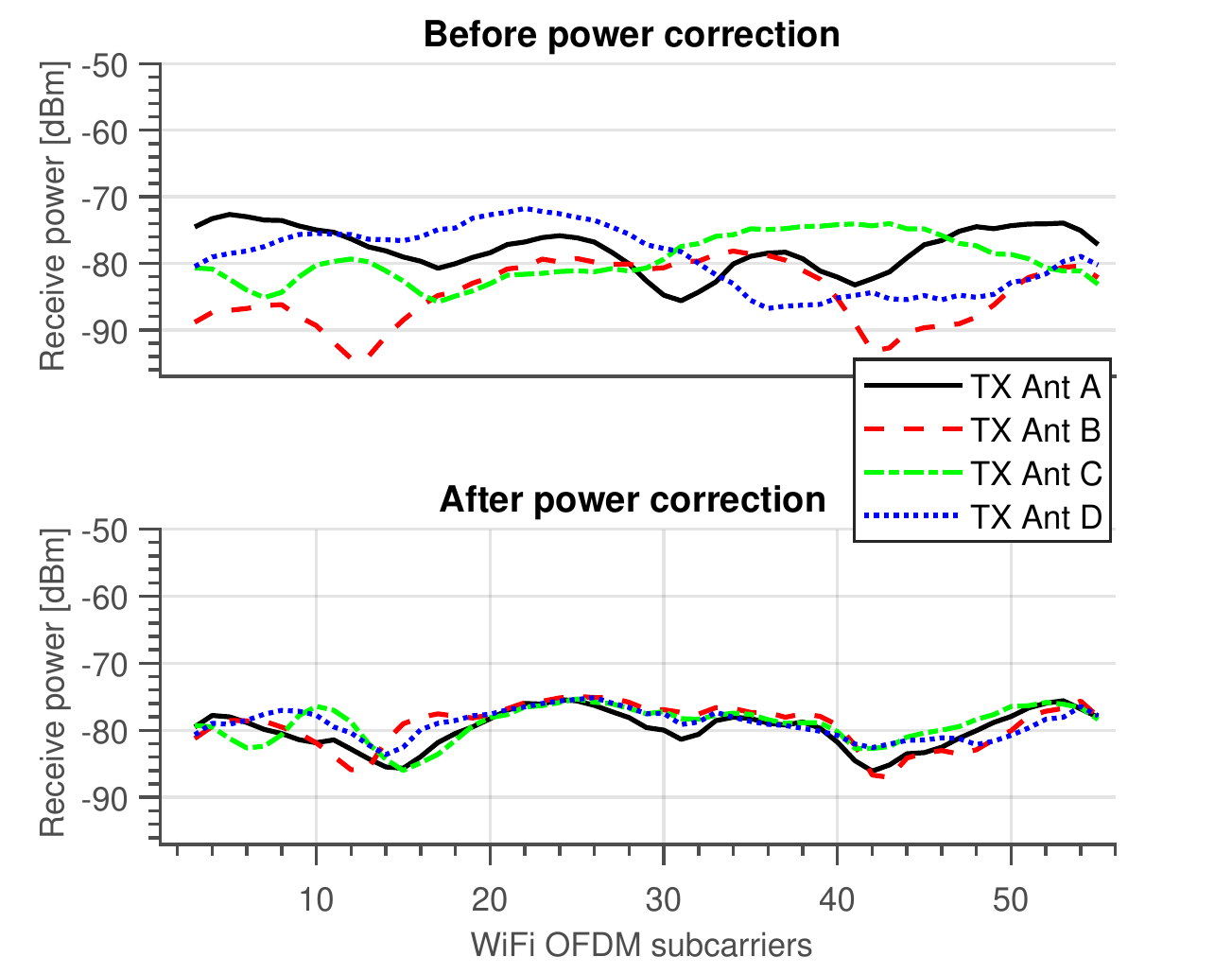}
\end{center}
 \vspace{-10pt}
\caption{Receive power of each antenna path at node13-13 before and after power correction respectively.}
\label{fig:orbit_exp_node13_13_siso}
\end{minipage}
\vspace{-15pt}
\end{figure}

While the interference reduction achieved by \methodname~in the previous scenario is very promising, unfortunately such a high reduction in INR may not be achievable in the existence of multipath effects that would occur in a real wireless channel.
To evaluate the robustness of \methodname~against multipath effects, we perform an over-the-air experiment under real conditions with severe multipath~(i.e. frequency selective fading) channel. 
As an example, Fig.~\ref{fig:orbit_exp_node13_13_siso} (upper) shows the received power at a randomly selected WiFi node from each BS's TX antenna.
We see very strong frequency selectivity of up to 12\,dB, which may hamper the operation of \methodname~as interference nulling becomes difficult in such a channel. 
Hence, in \methodname~we perform power correction (Sec.~\ref{sec:power_corr}).
The receiver power after correction is shown in Fig.~\ref{fig:orbit_exp_node13_13_siso} (lower).
Here, all antenna paths have similar receive signal strength and the total emitted power over all TX antennas stays the same after power correction, i.e., some paths are strengthened others weakened.

\begin{figure}[!t]
\centering
  \begin{tabular}{@{}c@{}}
    \includegraphics[width=0.75\linewidth]{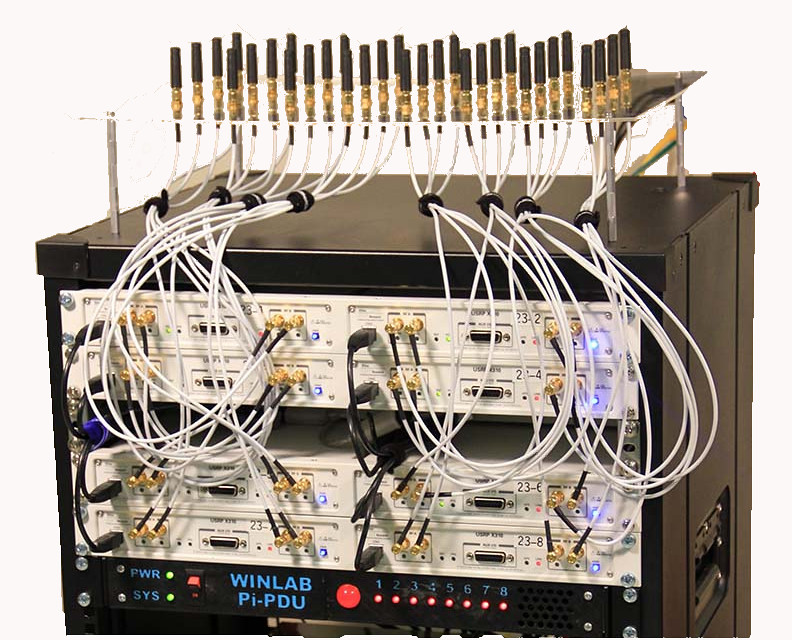}   \\
    \includegraphics[width=1\linewidth]{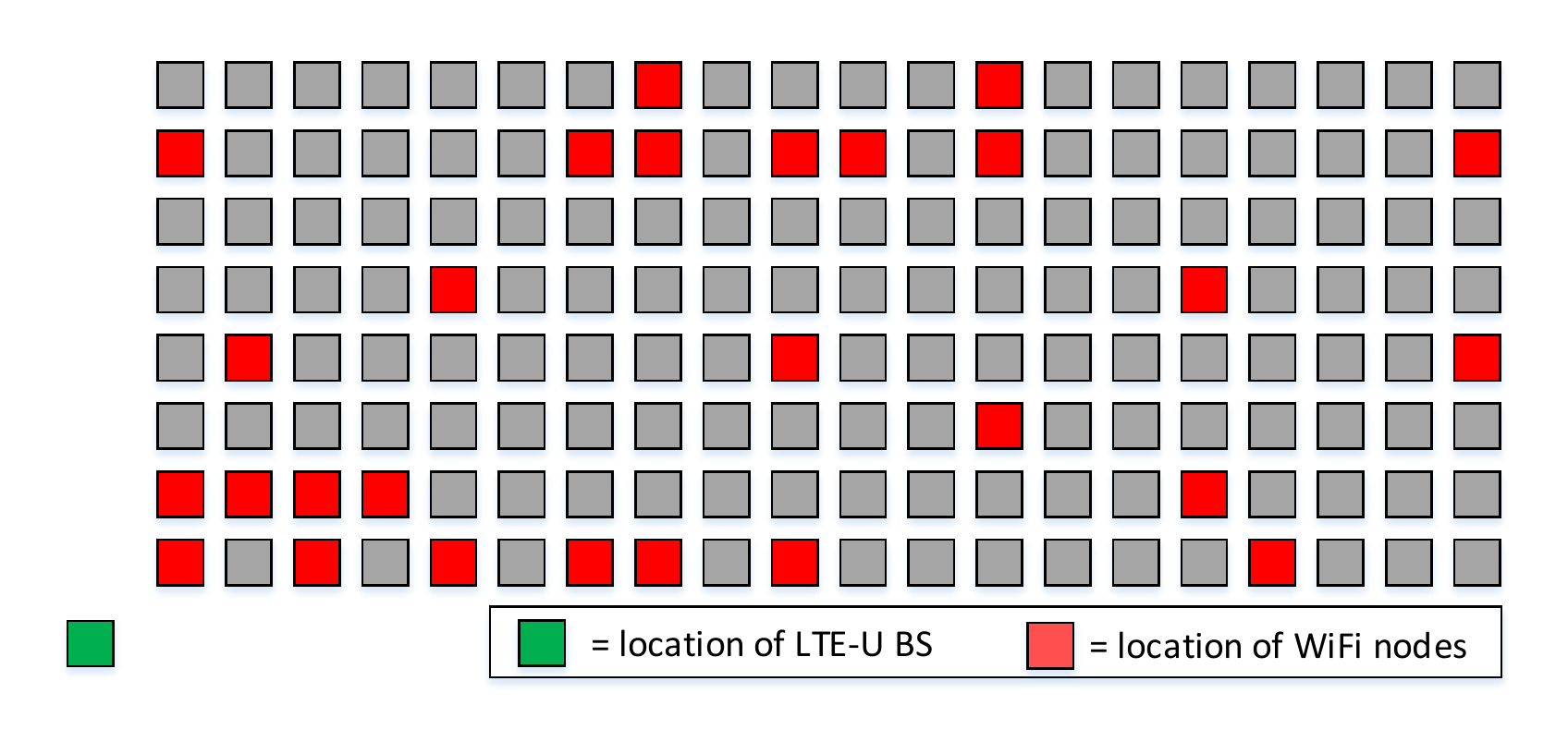}
  \end{tabular}
  \caption{Experiments in ORBIT grid --- mMIMO mini-rack used for \methodname~transmitter (upper) and node placement in grid layout with $\approx 1$\,m spacing (lower).}
  \label{fig:orbit_grid} 
  \vspace{-5pt}
\end{figure}

The BS's transmitter hardware used during this experiment is shown in Fig.~\ref{fig:orbit_grid} (upper). We selected $K=4$ TX antennas arranged along a line (ULA) with spacing of 7.18\,cm. The RF center frequency was selected as 2.412\,GHz (WiFi channel 1 in ISM band) as the antenna spacing was fixed in the ORBIT grid and too large for 5\,GHz UNII band.

\begin{figure*}[!t]
\centering
\begin{minipage}[b]{0.9\linewidth}
\begin{center}
    \includegraphics[width=1\linewidth]{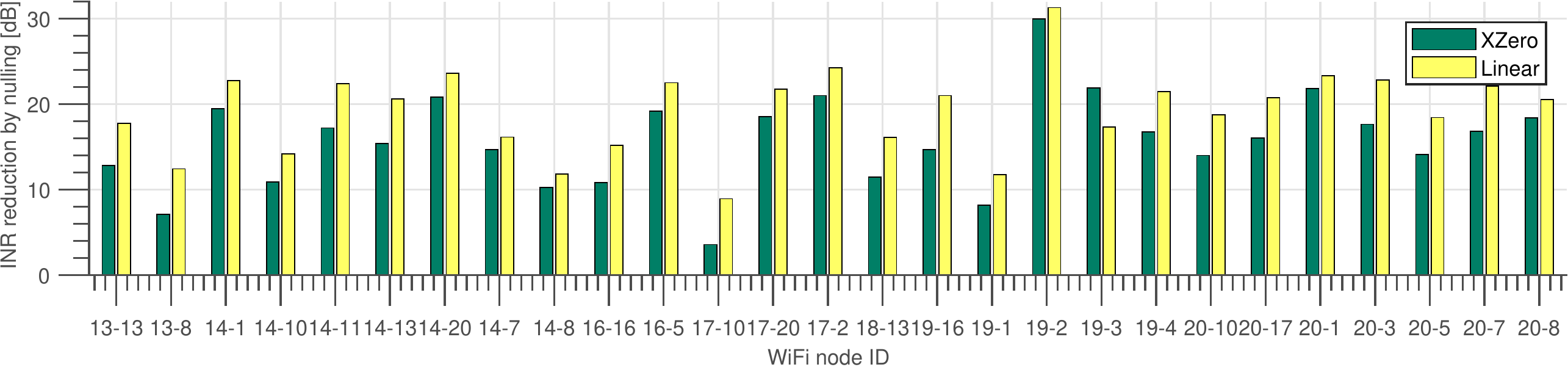}
\end{center}
 \vspace{-10pt}
\caption{INR reduction after nulling for both \methodname~'s tree-search and exhaustive linear search.}
\label{fig:orbit_exp_inr}
\end{minipage}
\vspace{-15pt}
\end{figure*}

For the experiment, we randomly selected 27 WiFi nodes equipped with Atheros 802.11n NIC from the ORBIT grid~\footnote{\url{www.orbit-lab.org}}. The placement of the BS and the location of the WiFi nodes are shown in Fig.~\ref{fig:orbit_grid} (lower).
Next, we executed the two null search algorithms, namely \methodname's tree and linear search, and recorded the reduction in INR~($\Delta$INR) due to nulling as compared to baseline without nulling.
As Fig.~\ref{fig:orbit_exp_inr} shows, we observe significant reduction in INR for both schemes.
More specifically, the average $\Delta$INR for \methodname~is 15.7\,dB while for some nodes the INR reduction can be up to 30\,dB.
However, \methodname's tree-search achieves in general a slightly lower $\Delta$INR compared to that of linear search.
We attribute this difference to possible wrong decisions made during the tree search, i.e. wrong subtree traversed.
However, as we show in the next section, the reconfiguration delay of \methodname~is up to $10 \times$ lower than that of the linear search.
This results in a tradeoff between null search speed and achieved $\Delta$INR.


Moreover, in a wireless channel with strong multipath \methodname~places multiple nulls even for single users as it chooses the best configuration from the tested nulling configurations.
Hence, it is possible that in some situations, a nulling configuration from an inner node of the tree achieves better INR than those tested in the leaf nodes.
%
%
In our experiment \methodname~ uses 2.7 nulls on average for a single user, i.e. WiFi node, to be nulled.
Again, we have a tradeoff between null-beam search speed and the required number of nulls. 
%

\begin{figure}[!t]
\centering
\begin{minipage}[b]{0.9\linewidth}
\begin{center}
    \includegraphics[width=1\linewidth]{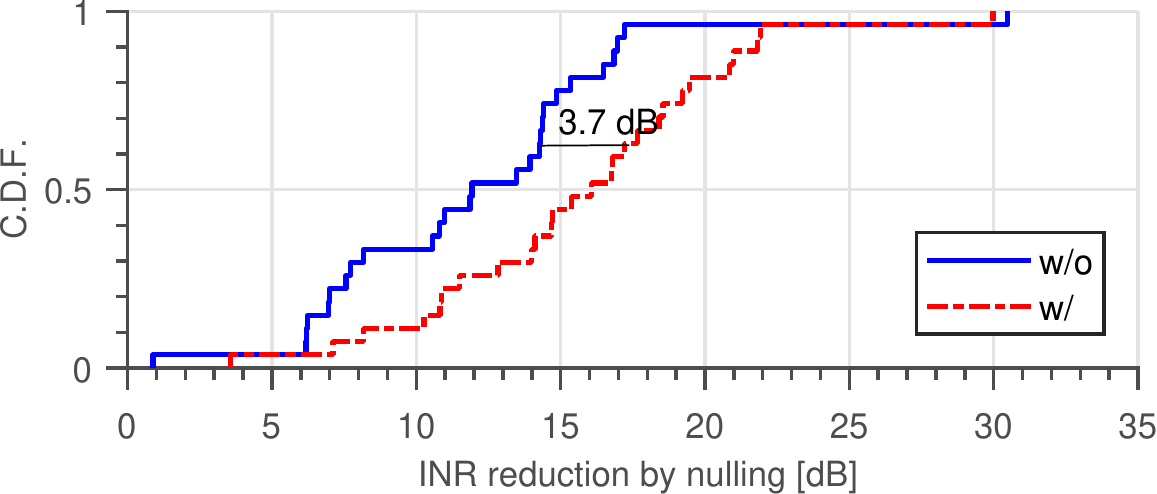}
\end{center}
 \vspace{-10pt}
\caption{INR reduction due to power correction.}
\label{fig:orbit_cmp_power_corr}
\end{minipage}
\vspace{-15pt}
\end{figure}

Finally, Fig.~\ref{fig:orbit_cmp_power_corr} shows the advantage from the proposed power correction (Sec.~\ref{sec:power_corr}). We see that on average 3.7\,dB is achieved compared to when power correction is not applied.

%
%
\subsection{Reconfiguration delay}
Null search has to be performed both at the time of network booting and also upon a change in network topology, e.g., due to node mobility as well as when WiFi nodes join or leave the network. 
\methodname~can support nomadic or even mobile WiFi nodes by reconfiguring the nulling configuration. 
Even when nodes are stationary, fast reconfiguration capability is desirable as the environment is not static, e.g. moving people.

As described in Sec.~\ref{sec:ctcbeam_smo}, the LTE-U BS tests different nulling configurations during \methodname's tree-based null search.
That is, during a single LTE-U CSAT cycle, $T_{\mathit{csat}}$, we can test multiple nulling configurations. This is because we are able to sample the receive signal power at a high sample rate of up to 50\,kHz at the WiFi receiver side~(Sec.~\ref{sec:proto}). From our experiments, we found out that for every 2\,ms of LTE-U $T_{\mathit{on}}$ phase a single nulling configuration can be tested with sufficient accuracy, i.e., averaging over 100 measured INR values.
Moreover, at each level of the search tree, the WiFi node informs the LTE-U BS about the best nulling configuration using the wired backbone so that the searching can continue in the proper subtree. Hence, both the LTE-U duty cycle, $\lambda_{\mathit{dc}}$, as well as the latency of the wired backhaul, $\delayBackhaul$, affect the reconfiguration delay.

%
%

\begin{figure}[!t]
\centering
\begin{minipage}[b]{1\linewidth}
\begin{center}
    \includegraphics[width=1\linewidth]{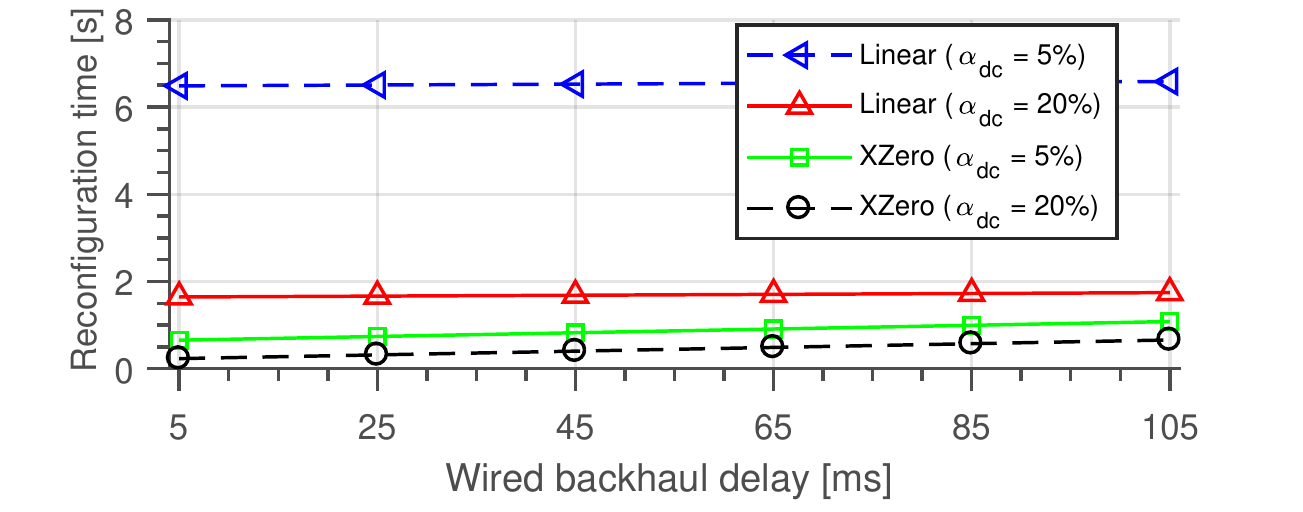}
\end{center}
 \vspace{-10pt}
\caption{Null search reconfiguration delay for a single WiFi user ($T_{\mathit{csat}}=40$\,ms).}
\label{fig:reconfig_delay}
\end{minipage}
\vspace{-15pt}
\end{figure}

Fig.~\ref{fig:reconfig_delay} shows the reconfiguration delay for different LTE-U duty cycles, i.e. 5 and 20\% respectively. The length of the CSAT period was fixed to $T_{\mathit{csat}} = 40$\,ms.
Corresponding $T_{\mathit{on}}$ duration is 2\,ms and 8\,ms for these two $\lambda_{\mathit{dc}}$ values, respectively.
Hence, during a single $T_{\mathit{csat}}$ period, \methodname~can test one and all three configurations~\footnote{In each LTE-U cycle, \methodname~can test at most three configurations as the search tree has a fanout of three. Hence, it cannot benefit from very long $T_{\mathit{on}}$. 
} for $\lambda_{\mathit{dc}}$ of 5\,\% and 20\,\%, respectively.
We can see that $\lambda_{\mathit{dc}}$ has a huge impact on linear search as with smaller value less nulling configurations can be tested during a single LTE-U cycle. In contrast, its impact is marginal in \methodname. 
Unsurprisingly, latency of the wired backhaul has a significant impact on \methodname~as the WiFi node sends its feedback over the wired backhaul for null search.
Nevertheless, the proposed tree-based null search can be completed in just 1.1\,s even under a high wired-backhaul latency of $\delayBackhaul=105$\,ms, i.e. Internet backhaul over different ISPs, and short LTE-U duty cycle of $\lambda_{\mathit{dc}}=5$\,\%.
This value is much faster compared to 6.6\,s needed for scanning the whole space, i.e., linear search.
For $\lambda_{\mathit{dc}}=20$\,\%, the tree search is completed in just 0.65\,s. The speed-up against linear search is $10 \times$.
Finally, under optimal conditions with a very fast backhaul of $\delayBackhaul=5$\,ms \methodname~requires 0.23\,s for reconfiguration.
We also observe that \methodname~is less sensitive to $\lambda_{\mathit{dc}}$ and hence to the network load in the LTE-U cell.

\begin{figure}[!t]
\centering
\begin{minipage}[b]{1\linewidth}
\begin{center}
    \includegraphics[width=1\linewidth]{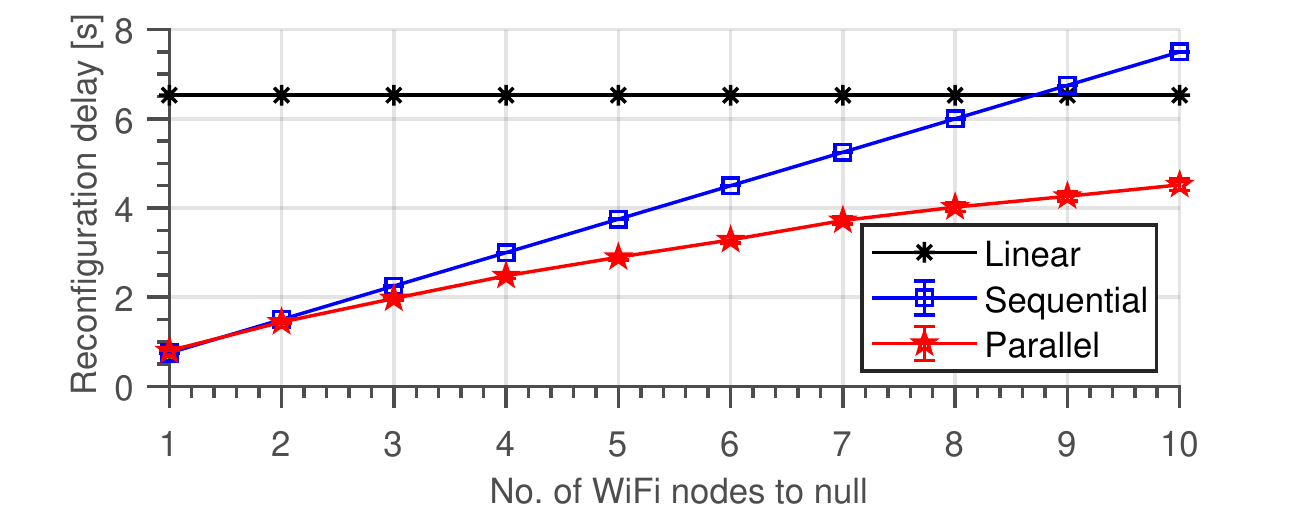}
\end{center}
 \vspace{-10pt}
\caption{Multi-user reconfiguration delay ($T_{\mathit{csat}}=40$\,ms, $\delayBackhaul=50$\,ms, $\lambda_{\mathit{dc}}=5$\,\%).}
\label{fig:cmp_parallel_search_ng}
\end{minipage}
\vspace{-15pt}
\end{figure}

Finally, Fig.~\ref{fig:cmp_parallel_search_ng} shows the delay for reconfiguration in case of multiple WiFi nodes to be nulled. We clearly see the speed-up due to the use of proposed parallel search as compared to sequentially executing the tree search for each user independently.

%
%
\subsection{Discussion}
The reconfiguration delay is very low with \methodname~at the cost of slightly increased INR and the usage of more than a single null.
Hence, in a mobile environment \methodname~should be preferred over exhaustive linear search. 
Given the fact that massive MIMO is becoming a reality, we expect the disadvantage of having to place multiple nulls per user to become insignificant. 
We have not detailed how to design the search tree. The tree's fanout, i.e. node degree, is one of the factors determining the reconfiguration latency. Another is the sampling frequency at the WiFi nodes which determines the time needed~($\tau_{s}$) for measuring INR under a nulling configuration. For a given $T_{on}$, \methodname~can test $\lfloor{T_{on}/\tau_{s}}\rfloor$ configurations. Then, \methodname~can adapt its tree search's fanout to fully utilize LTE-U on-period by setting it to $\lfloor{T_{on}/\tau_{s}}\rfloor$. Alternatively, it can apply speculative branching in which $\lfloor{T_{on}/\tau_{s}}\rfloor$ nodes are visited in each step of the tree search.


\section{Related Work}

\methodname~is a practical solution aiming to improve inter-technology coexistence between LTE-U and WiFi networks. 
While the literature on this general scope is very rich~(e.g., \cite{voicu_coexsurvey17}), practical solutions are only a few~\cite{yun2015supporting}. 
We identify two groups of research as the most related ones to \methodname: MIMO-based interference coordination and beam searching methods.

\medskip

\noindent \textbf{Interference Coordination:} %
The classical approach to enable coexistence is to avoid interference by means of spectrum isolation in either time, frequency, code, or space. 
In contrast, \methodname~belongs to the category of interference coordination, where multiple transmissions can co-exist without isolation~\cite{tse2005fundamentals}. 
%

\noindent \textit{\textbf{(a) Cross technology Interference:}} %
TIMO~\cite{gollakota2011clearing} uses interference nulling at the WiFi transmitter and cross-technology decoding at the WiFi receiver to enable cross-technology coexistence with other unlicensed technologies~(e.g., with wireless baby monitors or cordless phones). 
For nulling, a WiFi transmitter needs channel state information (CSI) to the colocated receiver of the other wireless technology which is obtained by utilizing channel reciprocity.
%
%
To achieve robustness in channel estimation, TIMO needs to sample the interferer's signal for a few seconds, which makes it difficult to apply in mobile settings.
To tackle the same challenge, \methodname~performs a quick null search 
rather than trying to estimate the channel between the two network nodes~(i.e., LTE-U BS and WiFi node) directly. 
TIMO requires substantial changes to the WiFi receivers to facilitate cross-technology signal decoding 
under strong interference. Moreover, the WiFi nodes have to be equipped with at least two antennas. 
On the contrary, \methodname~achieves nulling much faster compared to TIMO and without any significant changes to neither to LTE UEs nor to WiFi nodes. 
Moreover, \methodname~can operate under moderate node mobility.
On the other hand, our desire to keep the receivers unmodified may result in interference at the UEs from nulled WiFi nodes. 
While \methodname~can avoid this challenge by carefully selecting the WiFi nodes to be nulled, e.g., they must be distant and angularly separated from UEs, we plan to enhance \methodname~by implementing some receiver-side solutions in future work. 
Nevertheless, \methodname~and TIMO can complement each other: the former being implemented at the LTE-U BS and the latter at the WiFi AP.

Yun et al.~\cite{yun2015supporting} were first to consider cross-technology MIMO for supporting LTE and WiFi coexistence. Similar to TIMO~\cite{gollakota2011clearing}, they proposed a decoding scheme where LTE and WiFi transmitters are sending together and the receivers equipped with multiple antennas decode the overlapping (interfering) OFDM transmissions. However, they assumed the extreme case where LTE is transmitting continuous so that a special algorithm is needed to obtain the cross-technology channel state without the need to estimate a clean reference signal.
This is not needed as LTE-U uses duty cycled channel access.
Finally, the approach from~\cite{yun2015supporting} shares the same disadvantages with TIMO, e.g. modifications needed at receiver side (LTE-UE and WiFi STA) for complex signal processing.

\noindent \textit{\textbf{(a) Intra-technology Interference:}} %
OpenRF~\cite{kumar2013bringing} shares in essence the same goals as \methodname: utilizing MIMO interference nulling to facilitate better sharing of the available radio spectrum among co-located wireless nodes and networks.
In contrast to \methodname, OpenRF does not have to tackle the challenges of cross-technology communication as it is a solution for managing intra-network interference of Enterprise 802.11 WiFi networks.
%
%
%
We believe that \methodname~and OpenRF can complement each other as the former aims to coordinate the coexistence of heterogeneous technologies, i.e. mitigation of inter-technology interference, and the latter of homogeneous technologies, i.e. mitigation of intra-technology interference.

While the focus of OpenRF is on the downlink, BBN~\cite{zhou2014bbn} aims to improve the uplink reception by enabling multiple Enterprise WiFi APs to receive uplink traffic concurrently from their single-antenna clients. 
Rather than handling the interference nulling and beam forming at the client side, 
BBN exploits the wired backbone for communication between APs to minimize the complexity of BBN at the client side.
Through the backbone, APs share the decoded packets strategically to help each other extract the intended traffic from their respective clients, in other words null the unintended signals.
Other similar approaches using MIMO transmissions within the same technology are \cite{aryafar2010design,tan2009sam}.

%
%
%


In short, while \methodname~shares some goals and design principles with the above-listed MIMO interference coordination schemes, it is unique as it can be easily integrated with both the LTE-U and IEEE 802.11 standards. Moreover, we showed that receiver side can be implemented using existing commodity WiFi hardware.

%
%


\medskip

\noindent \textbf{Beam Searching Methods:} 
Agile-Link~\cite{abari2016millimeter} aims to find the correct beam alignment for an mmWave link which is frequently needed at the BS due to disruptions by beam misalignment, node mobility, or link blockage~\cite{patra_wowmom17}. 
Similar to \methodname’s tree search, Agile-Link applies a multi-step steer search instead of sequentially testing all beam directions to decrease beam-alignment time, e.g., typically multiple seconds. More specifically, Agile-Link eliminates the bins which do not accommodate a high-energy level. 
\methodname~has the same spirit as Agile-Link, however, the way it achieves its goals differs from the latter, i.e. tree-based search vs. beam hashing.

\section{Conclusions}
In this paper, we have developed a practical system, \methodname, for  improving the coexistence between co-located LTE-U and WiFi networks. 
%
%
Developing coexistence schemes is nontrivial due to the difference between LTE and WiFi medium access rules, and the lack of coordination mechanisms between these two networks. 
Our solution \methodname~tackles the first challenge by exploiting multiple antennas at the LTE BS to suppress the interference at the WiFi nodes while continuing its downlink transmission, in addition to the existing duty-cycling in LTE-U.
For the latter challenge, \methodname~ extends an existing solution for cross-technology communication between LTE-U and WiFi networks.
To the best of our knowledge, \methodname~is the first of its kind: a low-complexity practical system performing cross-technology interference-nulling between LTE-U and WiFi networks by applying a tree search to find the proper nulling configurations via feedback from the WiFi network.
Experimental analysis on the developed prototype shows that \methodname~can achieve a significant interference reduction on the nulled WiFi nodes.
Moreover the null search is very fast compared to exhaustive linear search. 
However, this comes at the cost of nulling multiple angles simultaneously in an environment with multipath fading. Given that massive MIMO is becoming viable, we expect this limitation to become insignificant. 


\section*{Acknowledgment}
This work has been supported by the European Union's Horizon 2020 research and innovation program under grant agreement No 645274~(WiSHFUL project).
\bibliographystyle{IEEEtran}
\bibliography{biblio}
\end{document}